\documentclass[11pt]{article}
\usepackage{amsmath,amssymb,amsthm,mathtools,geometry,booktabs,hyperref}
\usepackage{microtype}
\usepackage{natbib}

\title{\textbf{Computation as a Game}}
\author{Paul Alexander Bilokon\thanks{Department of Mathematics, Imperial College London, South Kensington Campus, London SW7~2AZ. Email: \texttt{paul.bilokon@imperial.ac.uk}}}
\date{28 October 2025}

\theoremstyle{plain}
\newtheorem{definition}{Definition}[section]
\newtheorem{theorem}[definition]{Theorem}

\theoremstyle{remark}

\begin{document}
\maketitle

\begin{abstract}
We present a unifying representation of computation as a two-player game between an \emph{Algorithm} and \emph{Nature}, grounded in domain theory and game theory. The Algorithm produces progressively refined approximations within a Scott domain, while Nature assigns penalties proportional to their distance from the true value. Correctness corresponds to equilibrium in the limit of refinement. This framework allows us to define complexity classes game-theoretically, characterizing $\mathbf{P}$, $\mathbf{NP}$, and related classes as sets of problems admitting particular equilibria. The open question $\mathbf{P} \stackrel{?}{=} \mathbf{NP}$ becomes a problem about the equivalence of Nash equilibria under differing informational and temporal constraints.

We then extend the game-theoretic interpretation of computation into a probabilistic and collective setting, formulating a \emph{mean-field theory of computation}. Each individual computation is modeled as a stochastic process on a domain, interacting weakly with the aggregate distribution of other computations.
We derive equilibrium conditions analogous to those of mean-field games and interpret complexity classes in terms of the structure and stability of these equilibria.
The $\mathbf{P}$ vs.\ $\mathbf{NP}$ question thus becomes a question of whether deterministic mean-field equilibria can approximate their existentially quantified counterparts in the large-population limit.

Finally, we present a categorical formulation of the game-theoretic theory of computation.
In this view, programs are morphisms in a cartesian closed category, computational refinement corresponds to the internal order in a domain object, and equilibria are fixed points of endofunctors representing interaction between an Algorithm and Nature.
Complexity classes arise as subcategories defined by resource-bounded morphisms.
The question $\mathbf{P}\stackrel{?}{=}\mathbf{NP}$ becomes the question of whether two endofunctors---deterministic and existential---are isomorphic in the topos of feasible computations.
\end{abstract}

\section{Introduction}

The classical Church–Turing thesis characterizes computation as mechanical symbol manipulation. Domain theory \citep{Scott1970, GierzEtAl1980, Edalat1993DSMF, AbramskyJung1994} recasts this as continuous approximation: each computation is a monotone function on a partially ordered set of information states. Game semantics \citep{AbramskyJagadeesanMalacaria2000} provides a complementary perspective, representing computation as an interaction between a \emph{Prover} and an \emph{Environment}. 

We unify these perspectives through a game-theoretic formalism in which computation is a \emph{game of approximation}. Each move refines partial information, and payoffs depend on both correctness and efficiency. We then reinterpret classical complexity theory through this lens: polynomial-time computability corresponds to the existence of equilibria under bounded strategic depth, while $\mathbf{NP}$ corresponds to equilibria under existential quantification over verification strategies.

\section{Computation as a Game: A Domain-Theoretic and Game-Theoretic Reformulation of Complexity Classes}

\subsection{Domain-Theoretic Representation}

Let $(D, \sqsubseteq)$ be a directed complete partial order (dcpo), representing the \emph{domain of computation}. For $x, y \in D$, we interpret $x \sqsubseteq y$ as ``$x$ is less informative than $y$''.

A \emph{computation} is a continuous map $f : D \to D$, such that
\[
f(\bigsqcup_i x_i) = \bigsqcup_i f(x_i)
\]
for any directed set $\{x_i\}$. Each stage of computation produces an approximation $f^n(\bot)$ approaching the fixed point $f^*(\bot)$.

\subsection{Computation as a Game}

We define a two-player game $G_f = (\mathcal{S}_A, \mathcal{S}_N, u_A, u_N)$:

\begin{itemize}
    \item $\mathcal{S}_A$ — strategy space of the \textbf{Algorithm} (choices of approximations $x_n \in D$);
    \item $\mathcal{S}_N$ — strategy space of \textbf{Nature} (assignment of penalties or rewards);
    \item $u_A(x_n, y)$ — utility to the Algorithm for producing approximation $x_n$ to true value $y$;
    \item $u_N = -u_A$.
\end{itemize}

Define the utility function as
\[
u_A(x, y) = -c(x) - d(x, y),
\]
where $c(x)$ is the computational cost of producing $x$, and $d(x, y)$ is a \emph{distance} in the metric completion of $D$, quantifying approximation error. The Algorithm seeks to maximize expected payoff (minimize cost and error), while Nature penalizes deviations from truth.

\begin{definition}[Computation Equilibrium]
A pair $(x^*, y^*)$ is a \emph{computation equilibrium} if
\[
x^* \in \arg\max_x u_A(x, y^*), \quad y^* \in \arg\min_y u_A(x^*, y).
\]
\end{definition}

This defines a Nash equilibrium of the computational game. Computation converges when iterative play leads to such an equilibrium.

\subsection{Complexity Classes as Equilibrium Families}

We define complexity classes via the structure of equilibria achievable under bounded resources.

\begin{definition}[Game-Theoretic Complexity Classes]
Let $C_T$ denote the class of games admitting an Algorithm strategy $x_n$ computable in time $T(n)$ such that the equilibrium distance satisfies $d(x_n, y^*) \le \varepsilon$ for some $\varepsilon \to 0$. Then:
\[
\mathbf{P} = \bigcup_{k \in \mathbb{N}} C_{n^k}, \qquad
\mathbf{NP} = \{ G_f : \exists \text{ verifier } V \text{ s.t. } (A,V) \text{ achieves equilibrium in polytime} \}.
\]
\end{definition}

Thus, $\mathbf{P}$ consists of games where equilibrium can be reached deterministically in polynomial time, whereas $\mathbf{NP}$ admits a \emph{witness-based equilibrium}: a second player (Verifier) ensures correctness in polynomial time given an existentially chosen proof.

\subsection{A Game-Theoretic Reformulation of $\mathbf{P} \stackrel{?}{=} \mathbf{NP}$}

Let $A_P$ be the Algorithm strategy space under polynomial-time constraints, and $A_{NP}$ include existential quantification over polynomial-time verifiers.

\begin{theorem}[Game-Theoretic P vs NP]
The question $\mathbf{P} \stackrel{?}{=} \mathbf{NP}$ is equivalent to asking whether
\[
\forall G_f \; \exists \text{ Nash equilibrium } (A_P, N) \quad \Leftrightarrow \quad
\forall G_f \; \exists \text{ Nash equilibrium } (A_{NP}, N),
\]
that is, whether deterministic equilibria coincide with existential equilibria in all computational games.
\end{theorem}

Intuitively, if for every computational game the deterministic player can achieve equilibrium without existential help, then $\mathbf{P} = \mathbf{NP}$; otherwise, $\mathbf{P} \neq \mathbf{NP}$ corresponds to the existence of games whose equilibria require verification strategies inaccessible to deterministic play.

\section{Computation as a Mean-Field Game:
Probabilistic Domain Semantics and the Complexity of Collective Computation}

\subsection{From Individual to Collective Computation}

A single computation $f:D\to D$ may be viewed as a game between an Algorithm and Nature.
When many such computations coexist—e.g., parallel threads, distributed agents, or neurons—the aggregate behaviour is best described by a distribution $\mu_t$ over $D$ evolving under collective dynamics.

Let $\mathcal{P}(D)$ denote the space of probability measures on $D$, equipped with the weak topology.
Each agent $i$ selects an approximation $x_i(t)\in D$ at time $t$ according to a strategy minimizing expected cost:
\[
J_i[x_i(\cdot),\mu(\cdot)] = 
\mathbb{E}\!\left[\int_0^T c(x_i(t),\mu_t) + d(x_i(t),y^*)\,dt\right],
\]
where $y^*$ is the true value and $\mu_t = \tfrac{1}{N}\sum_{i=1}^N \delta_{x_i(t)}$ is the empirical distribution.

\subsection{Mean-Field Limit and Equilibrium}

As $N\to\infty$, the empirical measure $\mu_t$ satisfies a deterministic Fokker–Planck equation
\[
\partial_t \mu_t + \nabla\!\cdot\!\big(\mu_t\,v(t,\mu_t)\big) = 0,
\]
where $v(t,\mu_t)$ is the best-response velocity field.
The optimal control $v$ solves the Hamilton–Jacobi–Bellman (HJB) equation
\[
-\partial_t \phi(t,x) = \min_{a}\big[ c(a,\mu_t) + d(a,y^*) + 
\nabla_x \phi(t,x)\cdot F(a,\mu_t) \big],
\]
with boundary condition $\phi(T,x) = d(x,y^*)$.
The pair $(\phi,\mu)$ constitutes a \emph{mean-field equilibrium} if both equations are satisfied self-consistently.

\begin{definition}[Mean-Field Computational Equilibrium]
A pair $(\phi,\mu)$ is a mean-field computational equilibrium if:
\begin{enumerate}
\item $\phi$ solves the HJB equation for given $\mu$;
\item $\mu$ solves the Fokker–Planck equation driven by the optimal control induced by $\phi$.
\end{enumerate}
\end{definition}

Intuitively, $\mu$ represents the distribution of partial computations across the population,
and $\phi$ encodes the ``value function''—the minimal expected cumulative penalty from each state.
At equilibrium, every computation’s expected improvement vanishes.

\subsection{Complexity as Energy Landscape}

The expected cost functional
\[
\mathcal{E}[\mu] = \int_D c(x,\mu)\,d\mu(x)
\]
defines an ``energy'' landscape on $\mathcal{P}(D)$.
We may interpret computational complexity as the curvature or metastability structure of this landscape:
steep basins correspond to rapid convergence (low complexity), while rugged landscapes encode combinatorial hardness.

\begin{definition}[Mean-Field Complexity Class]
Let $\mathcal{E}_T$ be the class of mean-field equilibria whose relaxation time $\tau(\mathcal{E})$ satisfies $\tau(\mathcal{E}) = O(T(n))$.
Then
\[
\mathbf{P} = \bigcup_{k} \mathcal{E}_{n^k}, \quad
\mathbf{NP} = \{ \mathcal{E} : \exists \text{ witness flow } \nu_t \text{ verifying equilibrium in } O(n^k) \}.
\]
\end{definition}

\subsection{Game-Theoretic Reformulation of $\mathbf{P} \stackrel{?}{=} \mathbf{NP}$}

In the mean-field picture, $\mathbf{P} = \mathbf{NP}$ would hold if every equilibrium distribution attainable through existential (verification) dynamics could also be approximated by deterministic (best-response) dynamics with the same relaxation time.

\begin{theorem}[Equilibrium Reformulation of $\mathbf{P} \stackrel{?}{=} \mathbf{NP}$]
Let $\mathcal{E}_P$ denote deterministic mean-field equilibria and $\mathcal{E}_{NP}$ those achieved via stochastic existential verification.
Then
\[
\mathbf{P} = \mathbf{NP} \quad \Leftrightarrow \quad
\forall \mathcal{E}_{NP}\ \exists \mathcal{E}_P\ \text{with}\ W_1(\mathcal{E}_P,\mathcal{E}_{NP}) \le \varepsilon,
\]
where $W_1$ is the Wasserstein distance on $\mathcal{P}(D)$ and $\varepsilon \to 0$ polynomially.
\end{theorem}

Thus $\mathbf{P}\neq\mathbf{NP}$ would correspond to the existence of existential equilibria unreachable by deterministic computation within polynomial-time mixing.

\subsection{Connections to Learning and Stochastic Control}

The above formalism naturally connects with:
\begin{itemize}
\item \textbf{Reinforcement learning:} the HJB–Fokker–Planck pair is analogous to the policy–distribution update system in mean-field reinforcement learning.
\item \textbf{Statistical physics:} $\mathcal{E}[\mu]$ acts as a free energy functional, linking computational hardness to phase transitions.
\item \textbf{Adversarial optimization:} Nature’s penalty function corresponds to an adversary controlling environmental uncertainty.
\end{itemize}

\section{Computation as a Categorical Game:
A Topos-Theoretic Reconstruction of Complexity and Equilibrium}

\subsection{Computation as Morphism}

Let $\mathcal{C}$ be a cartesian closed category (CCC) with objects representing domains of data and morphisms representing computable transformations.  
For objects $A,B\in\mathrm{Ob}(\mathcal{C})$, a computation is a morphism $f:A\to B$.

Partiality and approximation are represented by an internal order $\sqsubseteq$ in $\mathcal{C}$:  
\[
x\sqsubseteq y \;\Leftrightarrow\; \exists m:A\to B \text{ s.t. } m(x)=y .
\]
Each morphism $f$ induces a monotone endomap $T_f:B^A\to B^A$, and computation corresponds to the search for a fixed point $f^\ast$ of $T_f$.

\subsection{Game Functors}

Define two endofunctors on $\mathcal{C}$:
\[
\mathsf{Alg}(X)=A\times X,\qquad
\mathsf{Nat}(X)=B^X,
\]
interpreted respectively as the Algorithm’s action and Nature’s reaction.
Their interaction is described by the composite endofunctor
\[
\mathsf{Comp}(X)=\mathsf{Nat}(\mathsf{Alg}(X))=B^{A\times X}.
\]
A \emph{computational game} is then a coalgebra $(X,\gamma)$ for $\mathsf{Comp}$, where $\gamma:X\to B^{A\times X}$ encodes the state-transition structure of the computation.

\begin{definition}[Categorical Equilibrium]
A morphism $\eta:X\to B$ is an equilibrium if it is a \emph{coalgebra morphism}:
\[
\eta = \mathsf{Comp}(\eta)\circ \gamma.
\]
Equivalently, $\eta$ is a fixed point of $\mathsf{Comp}$ in the sense of Lambek’s lemma.
\end{definition}

Such equilibria generalise Nash equilibria: no natural transformation (strategy) can unilaterally increase its categorical utility.

\subsection{Complexity Classes as Subcategories}

Let $\mathcal{C}_{\le T}\subseteq\mathcal{C}$ be the full subcategory of morphisms computable within resource bound $T(n)$, e.g.\ polynomial time.
Then
\[
\mathbf{P} = \bigcup_k \mathcal{C}_{\le n^k}, \qquad
\mathbf{NP} = \exists\text{-closure}(\mathbf{P}),
\]
where the existential closure is the image under a left Kan extension functor $\exists:\mathcal{C}\to\mathcal{C}$ representing nondeterministic choice or witness extraction.

\begin{definition}[Deterministic and Existential Functors]
Let $\mathsf{F_P}$ denote the deterministic functor restricting to $\mathbf{P}$ and
$\mathsf{F_{NP}}=\exists\!\circ\!\mathsf{F_P}$ the existential extension.
\end{definition}

\begin{theorem}[Categorical Form of $\mathbf{P}\stackrel{?}{=}\mathbf{NP}$]
The statement $\mathbf{P}=\mathbf{NP}$ holds if and only if
\[
\mathsf{F_P}\cong\mathsf{F_{NP}}
\]
as endofunctors in the topos $\mathcal{E}$ of feasible computations.
\end{theorem}

Thus $\mathbf{P}\neq\mathbf{NP}$ corresponds to a broken natural isomorphism:  
there exists no natural transformation $\theta:\mathsf{F_P}\Rightarrow\mathsf{F_{NP}}$
with a polynomially computable inverse.

\subsection{Topos of Feasible Computations}

Construct a topos $\mathcal{E}$ whose objects are resource-bounded domains and morphisms are feasible (polynomial-time) arrows.
Sheaves over $\mathcal{E}$ represent families of computations varying over contexts (inputs, resources).

The internal logic of $\mathcal{E}$ captures computational provability:
\[
\mathbf{P}\text{-computable} \;\Leftrightarrow\; \mathcal{E}\vDash \exists f:A\to B,\ \mathrm{polytime}(f).
\]
Existential quantification in this internal logic models nondeterministic verification.
Hence the $\mathbf{P}$ vs $\mathbf{NP}$ problem translates to whether this topos is Boolean (deterministic completeness) or merely Heyting (constructive incompleteness).

\section{Conclusion}

The domain-theoretic and game-theoretic formalism connects complexity, approximation, and rational behaviour in a single framework. In particular:
\begin{itemize}
    \item Domain theory captures the structure of partial information.
    \item Game theory quantifies the interactive process of refinement and verification.
    \item Complexity theory classifies the equilibria achievable under resource constraints.
\end{itemize}

Future work may explore \emph{quantitative domains} in the sense of Lawvere metrics, defining energy-based or probabilistic payoffs, and connecting to mean-field game limits for large-scale distributed computation.

The mean-field view of computation suggests that complexity classes encode collective equilibrium phenomena.
The $\mathbf{P}$ vs.\ $\mathbf{NP}$ problem expresses whether stochastic collective equilibria can be efficiently emulated by deterministic dynamics.

This interpretation links computational theory, game theory, and statistical physics within a single mathematical framework.

Stepping up the formalism, we obtain:
\begin{itemize}
\item \textbf{Domain theory:} computation as approximation sequence.
\item \textbf{Game theory:} computation as interactive optimisation.
\item \textbf{Category theory:} computation as a fixed point of functors.
\end{itemize}
Complexity arises as the geometry of morphisms within the topos of feasible computations,
and $\mathbf{P}\stackrel{?}{=}\mathbf{NP}$ becomes an isomorphism question in categorical semantics.

\section*{Acknowledgements}

The author thanks Samson Abramsky, Abbas Edalat, Claire Jones, Achim Jung, Pierre-Louis Lions, and Gordon Plotkin for foundational work on domain and game semantics, control, and category theory, and Imperial College London for fostering interdisciplinary research in computation and strategy.

\nocite{*}
\bibliographystyle{apalike}
%\bibliography{references}

\end{document}